# Real-world application of deep learning in large-scale seismic interference attenuation: A case study in the Camie field of Angola

Jing Sun[1,*], Song Hou[2]

[1]Delft University of Technology, Faculty of Electrical Engineering, Mathematics and Computer Science, Van Mourik Broekmanweg 6, Delft 2628 XE, The Netherlands.

[2] VIRIDIEN (UK) Ltd, Crompton Way, Manor Royal Estate, Crawley, West Sussex RH10 9QN, UK.

*Corresponding author: Jing Sun (jing.sun@tudelft.nl)

## ABSTRACT

In marine seismic acquisition, seismic interference (SI) occurs when energy from nearby external seismic source(s) is captured. It typically appears as coherent noise with linear or non-linear movement and varying amplitudes across different sail lines. SI is commonly observed and poses a challenge for seismic data processing. We present a case history of a previously proposed deep neural network (DNN)-based workflow applied for SI attenuation across a marine seismic block in the Camie Field of Angola. This field survey covers over 345 km$^2$ and is marked by the challenge of multiple SI types. The employed DNN-based workflow performs SI attenuation in the common shot domain based on a supervised learning framework: a small subset of the SI-contaminated data was first processed by a conventional geophysical algorithm to obtain an estimate of the SI noise, which was then manually blended with the SI-free common shot gathers from the same survey to generate the training pairs. To ensure signal fidelity, several techniques were applied to improve the DNN's performance. A key highlight of this case history is its scale: this represents a real-world, large-scale processing project and we present a comprehensive comparison of the DNN-based workflow with the conventional geophysical algorithm across the entire survey block, focusing on both processing quality and processing time. The results demonstrate the outstanding performance of the employed DNN-based workflow, which achieved higher SI removal accuracy, with less signal leakage and more complete SI removal. The promising results of this application also open up possibilities for integrating deep learning into other seismic denoising tasks. In addition, we discuss the limitations of this case history, aiming to provide insights for future research and applications in the field.

## INTRODUCTION

In densely worked offshore areas, seismic data are often contaminated by energy from nearby external seismic sources, typically air guns used by other vessels conducting seismic acquisition (Akbulut et al., 1984). This kind of noise is termed seismic interference, SI. SI presents as coherent linear or non-linear (curved) events within the shot domain, depending on the location of the external seismic source that generates it. Within a single survey, the angles of incidence for SI can vary significantly. Such variation is particularly noticeable when comparing one sail line to another, and it largely depends on the relative orientation of the external source in relation to the receivers. Similarly, notable differences in the amplitudes of SI can usually be observed across a survey, depending on the relative distance between the external seismic source and the receivers.

Originating from robust sources specifically designed for seismic exploration activities, SI often has a higher amplitude than other types of noise unless its seismic sources are very far away. In addition, SI propagates as guided waves traveling through the water column (Pekeris, 1948), which also contributes to its distinct presence. Another notable aspect of SI is its preservation capacity. SI can remain remarkably consistent over large distances (Akbulut et al., 1984; Jansen et al., 2013). A point of concern arises when this high-amplitude SI starts to overlap with desired seismic signals. These desired signals, which are essentially reflections from subsurface layers, tend to have a much lower amplitude. This overlapping can pose significant challenges in separating SI from seismic signals.

Based on the positioning of the external source in relation to the receivers, SI can be categorized into three classes: SI coming from ahead, SI coming from broadside and SI coming from astern of the recording vessel. Figure 1 provides a graphical representation of shot gathers contaminated by SI at varying directions and distances.

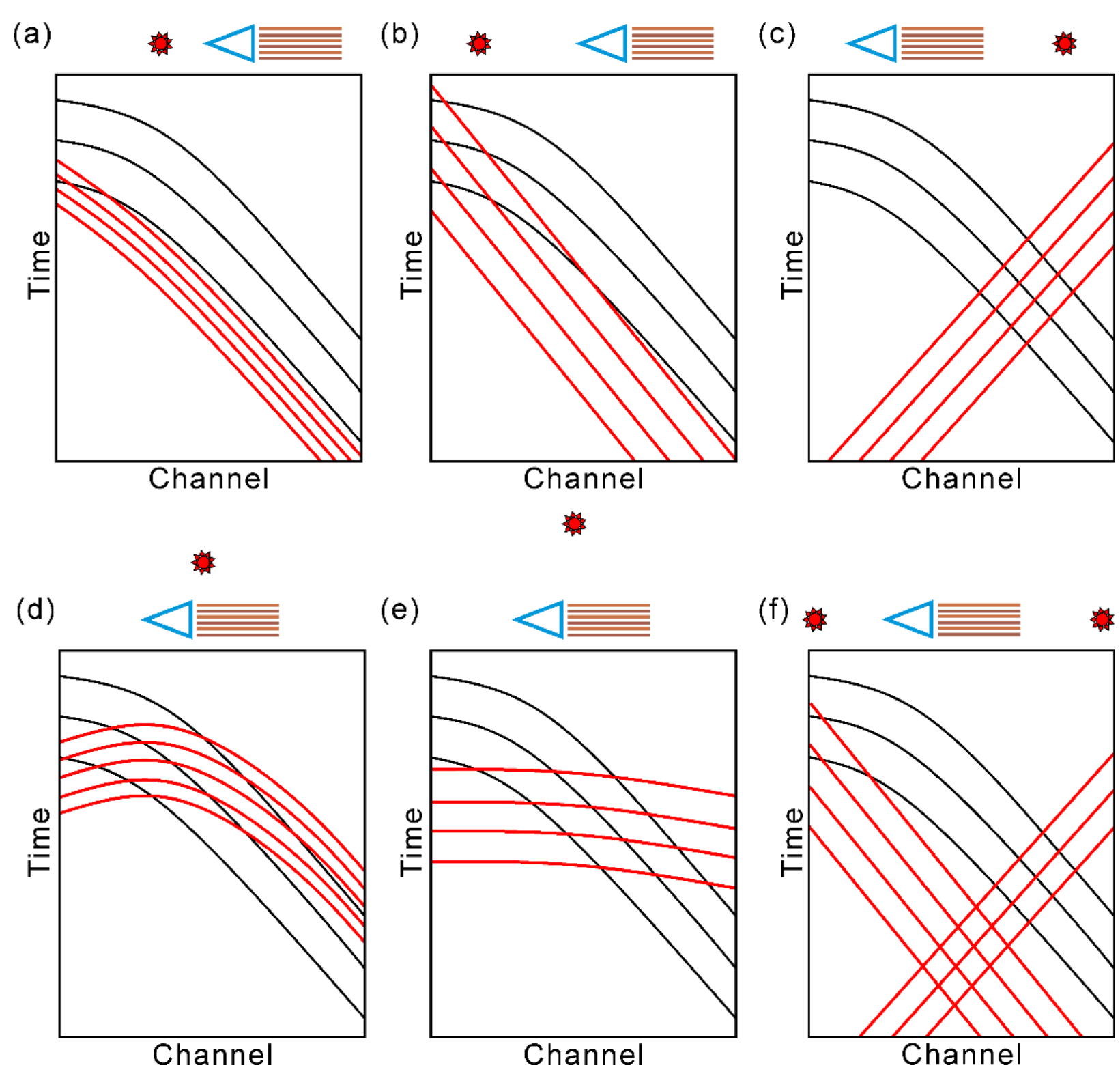

Figure 1: Schematics of shot gathers contaminated by SI from external source(s) at different positions. The blue triangle represents the recording vessel with streamers towed, and the red

spot represents the external seismic source. Black lines show seismic reflections from the linked source, while red lines indicate SI (adapted from Hlebnikov et al., 2021).

Additionally, due to the intensity of modern marine activities, SI often coexists with other forms of marine noise, such as noise from the mechanical operations of non-seismic vessels, mainly fishing and cargo vessels (Brittan et al., 2008). Figures 2a and 2b show typical traffic in two marine regions where seismic acquisitions are conducted, specifically west of Norway and Angola (where the studied survey block is located). The colored dots and arrows denote various vessels trackable by the automatic identification system (AIS). Considering untraceable vessels, actual marine traffic is likely even more congested than illustrated in the figures.

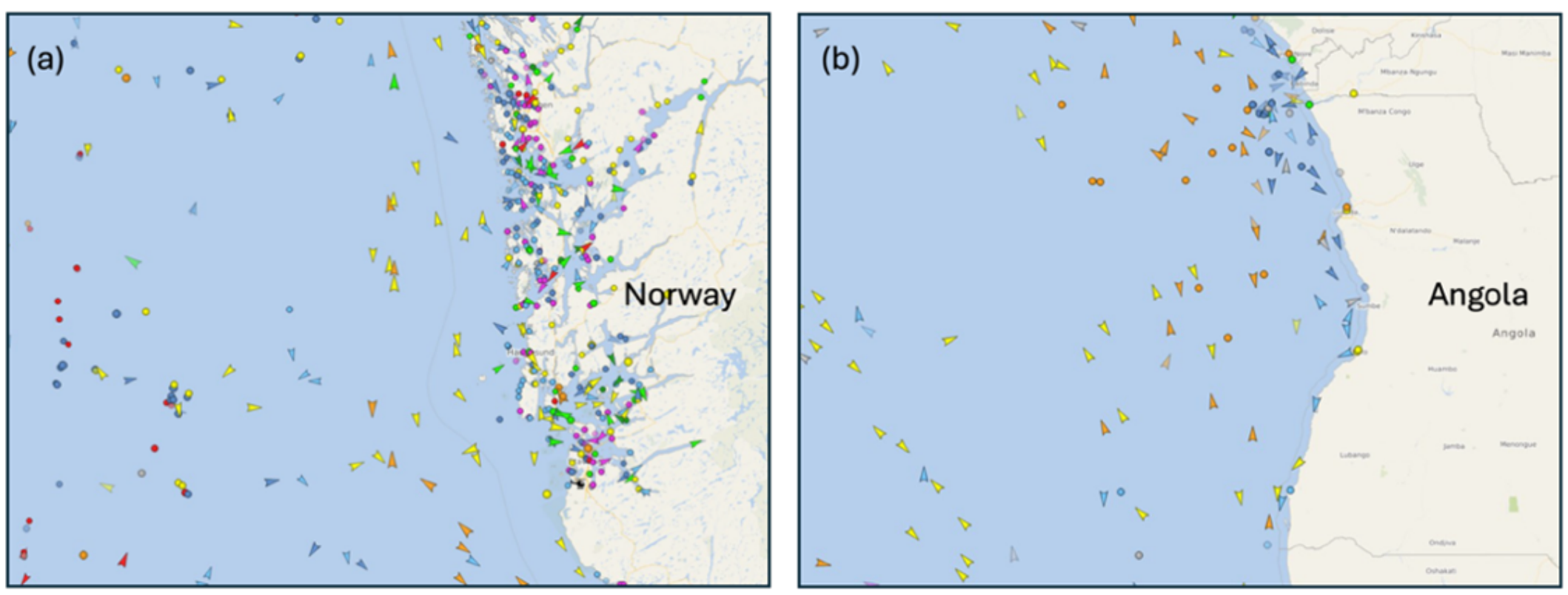

Figure 2: Examples of typical traffic in the marine seismic acquisition area: (a) and (b) show oceans to the west of Norway and Angola, respectively. Colored dots and arrows represent vessels for various uses monitored by the AIS.

Time-sharing was once considered a solution but proved inefficient, leading to unaffordable costs and vessel downtime (Elboth and Haouam, 2015). As a result, evolving strategies, especially cross-party coordination (Elboth 2017; Firth 2017) since 2015, have promoted synchronized vessel operations for effective SI control. The key idea is to accept the presence of SI types that can be attenuated at the processing stage and avoid including the highly intractable SI type that cannot be effectively attenuated, primarily those coming from close broadside and mirroring the moveout of desired seismic reflection data (Dhelie et al., 2013; Elboth and Haouam, 2015; Laurain et al., 2016). This enables a balance between acquisition time consumption and processing costs. Achieving this balance has been a dynamic process, as the tolerance for SI inclusion during acquisition depends on the knowledge of SI characteristics and advancements in SI attenuation techniques.

Over the years, many geophysical methods for the purpose of SI attenuation have been proposed. In general, they can be classified into two groups (Jansen, 2013). The first group involves breaking the coherence of SI by resorting the data into the common-receiver or common-offset domain, followed by the application of a denoising method such as f-x prediction filtering (Canales, 1984; Wang et al., 1989). Building on this, Gulunay and Pattberg (2001a, b) proposed guiding the detection of SI-contaminated shots through a magnitude threshold in the f-x-y domain, followed by the prediction and subtraction of inline coherent SI using very short f-x prediction filters, as well as applying an f-x-y prediction filter to the frequency slice. Guo and Lin (2003) similarly used inline f-x prediction filters but explored adaptive rather than straightforward subtraction. Gulunay et al. (2004) proposed an f-x-y domain method combining inline and crossline f-x prediction filters for detecting and attenuating strong SI noise, particularly effective at later record times with high SI amplitudes. Rajput and Rajput (2006) further studied the effectiveness of f-x-y prediction filters in 3D marine seismic data, demonstrating strong SI attenuation with good signal preservation. Kommedal et al. (2007) studied SI attenuation in multicomponent 4D data, particularly for permanently installed arrays.

In addition to the variants of f-x prediction filtering, Tau-P transform has also been commonly adopted for SI attenuation in the seismic industry as it separates SI noise from signals based on their different move-out behavior (dip and/or curvature) in the time domain (Elboth et al., 2010). Yu (2011) investigated SI attenuation by transforming data to the Tau-ρ domain and applying a multidomain 3D spatial filter based on amplitude and frequency attributes. Zhang and Wang (2015) improved this approach by using progressive sparse Tau-P inversion, which yielded better results than the conventional Tau-P transform in both synthetic and field examples.

Elboth et al. (2017) proposed a line-mixing method that combines two adjacent sail lines to identify and remove occasional shot-to-shot coherent SI, assuming SI differs significantly between the sail lines. However, since the mixing is done in a mirroring way, data recorded on the inner streamer cables may suffer from signal loss in areas with rapidly changing geology in the crossline direction. To address this, Shen et al. (2019) proposed an improved line-mixing method using a cascaded mixing algorithm to reduce signal loss. They also presented a shot-skipping and resplicing approach to decrease the consistency of shot-to-shot coherent SI.

The second group of SI attenuation methods is based on noise modelling and subtraction; the success of these methods strongly depends on their ability to build up an accurate model of the SI (Jansen 2013). Manin and Bonnot (1993) proposed that for noise independent of the recorded seismic shot, like SI, if the positions of the noise source and the receivers are known, the shape of the wavefront may be estimated and then refined using a shape recognition method when the noise train is long enough. Fookes et al. (2003) presented a similar method but estimated the noise source directly from the seismic data by identifying significant SI arrivals, enabling prediction of arrival times and flattening of noise trains via static shifts. Various methods, such as f-k or Radon transforms, median filters, or Karhunen-Loeve filters, can then be applied for denoising. Once the SI is attenuated, the data can be shifted back to their original time (Brittan et al., 2008). In practice, the data may be contaminated by SI from multiple sources and require iterative application of this method (Brittan et al., 2008).

Gulunay et al. (2005; 2006; 2008a) proposed the diffractor scan (DSCAN) method, which estimates noise source locations using semblance scan analysis on a surface grid. Gulunay (2008b) later proposed the Interfering Source Location Scanning (ISCAN) method, which also aims to remove SI from shallower depths. Guo et al. (2009) proposed another automatic technique that divides the shipping lane into a grid of cells to calculate a time trajectory for each, which requires prior knowledge of the lane, may be time-consuming and is limited in accuracy by the grid size. Lu et al. (2014) and Li et al. (2013) proposed an efficient method to calculate source locations from detected apices without scanning, though it fails when apices are not present on the seismic record. In a more recent work, Xu et al. (2019) applied density-based clustering to locate external SI sources by analyzing time delays between seismic traces, followed by iterative attenuation and an optional high-pass filter for detecting weak SI while preserving valid signals.

Although conventional geophysical SI attenuation methods have seen significant development, they still face important challenges. Removing SI noise with varying moveout (dip and/or curvature) across an entire survey typically requires manual testing and selection of different parameter sets, which can be labor-intensive. Moreover, in scenarios where SI noise originates from multiple directions or exhibits a dip similar to that of the underlying seismic reflections, conventional methods often struggle, resulting in noticeable residual noise and signal leakage.

Motivated by the great successful applications in computer vision and conventional image processing, in recent years, deep learning methods, especially various convolutional neural networks (CNNs), have emerged as potential alternatives to conventional seismic processing methods. Many related studies have been conducted within the seismic community on various processing tasks (Yu and Ma, 2021), such as deblending (Baardman and Tsingas, 2019; Richardson and Feller, 2019; Baardman and Hegge, 2020; Sun et al., 2020; Wang et al., 2020; Zu et al., 2020; Wang et al., 2021; Wang et al., 2023), ground roll attenuation (Li et al., 2018; Kaur et al., 2020; Yuan et al., 2020; Pham and Li, 2022; Yang et al., 2023), ocean bottom seismic noise attenuation (Wang et al., 2023; Seher et al., 2024; Wang et al., 2024; Zhu et al., 2025), as well as the focused problem of SI attenuation.

The earliest work on this topic can be traced back to Slang (2019), where the use of several CNN architectures was explored, laying the groundwork for using deep learning to attenuate SI. Slang et al. (2019) and Sun et al. (2020) further improved SI attenuation results in field marine seismic data. During the same period, Xu et al. (2020) also investigated DNN-based SI attenuation and proposed obtaining labeled training data through a sample generation method. While these earlier studies provided valuable insights and demonstrated the potential of DNNs for SI attenuation in field data, the processing quality had not yet been validated at a level consistent with industrial standards. Sun and Hou (2022) proposed a method to improve the DNN's signal fidelity for general seismic signal separation tasks and demonstrated the effectiveness of this method in attenuating SI. Building on this progress, Sun et al. (2023) developed a DNN-based workflow to attenuate coherent noise, and it was tested on attenuating SI in two sail lines of marine towed-streamer data acquired from the Northern Viking Graben area. However, the processing scale remained limited, and to date, no study has presented a full-block, large-scale application of DNN-based SI attenuation in direct comparison with a field-tested geophysical algorithm.

In this work, we present a much larger-scale real-world processing project of attenuating different types of SI using the DNN-based workflow proposed by Sun et al. (2023), adopting the same U-Net architecture as in their study, which was adapted from the U-Net introduced by Ronneberger et al. (2015). While previous applications of DNN-based SI attenuation have been restricted to smaller-scale or synthetic datasets, the present study encompasses an entire marine seismic block in the Camie Field of Angola, covering over 345 $km^2$. To the best of our knowledge, this represents the first published case study covering such a large area. As such, this research stands as a landmark in the field.

In the following sections, we introduce the employed geophysical algorithm and provide details on the implementation of the DNN-based workflow. The results include field data examples in the common shot domain and the stacked common midpoint (CMP) domain, along with detailed high-frequency root mean square (RMS) comparisons of individual sail lines and the entire survey. Based on the computational setup of the processing project, we also compare overall processing times of these two methods for the entire survey. The employed geophysical workflow follows a modular, multi-stage structure that is primarily CPU-based, with individual components optimized independently. This reflects a typical paradigm in commercial seismic processing systems. In contrast, the data preparation of the DNN-based workflow is CPU-based, while model training is executed on a graphics processing unit (GPU). We acknowledge that the comparison of processing times may not be as rigorous or directly comparable as in a pure research setup, where all algorithms are typically tested on the same graphics processing unit (GPU) or computational environment. However, this comparison aligns closely with most real-world production scenarios, where computational resources are often diverse, and optimization is tailored to specific project constraints. Finally, taking these aspects into consideration, we present a discussion and draw conclusions.

## THE STUDIED SURVEY

The acquisition area is in a deep-water environment off the coast of Angola, where broadband seismic 3D surveys were conducted (Cazier et al., 2014). The specific field we focus on is the Camie field. It is located in nearly 1700 meters of water and lies 100 kilometers offshore in Angolan Block 21, as shown in Figure 3 (Duval et al., 2015; Bhattacharya and Moore, 2026). This field includes 41 sail lines, each equipped with 12 cables. Each sail line has about 800 shots, which means that across the whole survey, there are about 400,000 (41x800x12) shot/cable combinations.

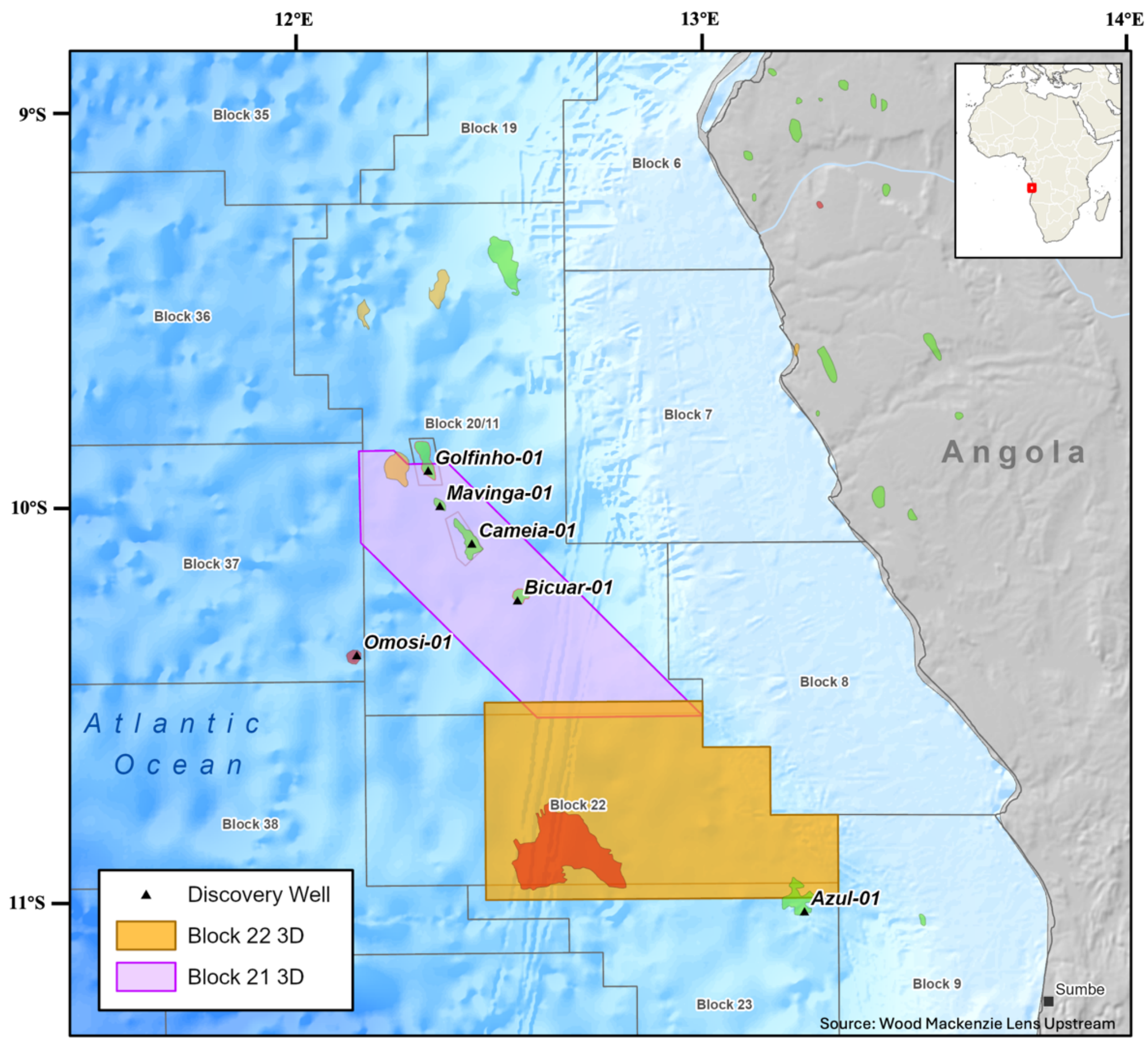

Figure 3: Location of the studied field in the Outer Kwanza Basin, Angola.

This acquisition took place during a particularly busy season, with SI noise caused by other seismic vessels conducting acquisition in the area, as well as coherent noise from numerous and unchecked local fishing vessels. Consequently, a wide range of SI, each with distinct characteristics, was recorded. Some of these SI types presented significant challenges in their removal from the dataset. For instance, in some cables, SI originating from two different directions was simultaneously captured within one shot gather, as shown in Figure 4a. Conventional parameter-based approaches then require repeated manual adjustments to suppress both types of SIand still risk incomplete suppression. In some other cables, the SI exhibited an extremely flat profile, as shown in Figure 4b. In addition, there was SI with a dip similar to that of the desired signals, as shown in Figure 4c. Such interference is difficult to distinguish from primary reflections using standard moveout- or dip-based filters, often resulting in residual noise or leakage of useful signals. Conversely, there were instances of SI that appeared to be the inverse of the former, as shown in Figure 4d. The diverse nature of these SI characteristics requested the development and application of advanced techniques, such as the deep learning approach used in this project, to effectively address these challenges and enhance the quality of the seismic data.

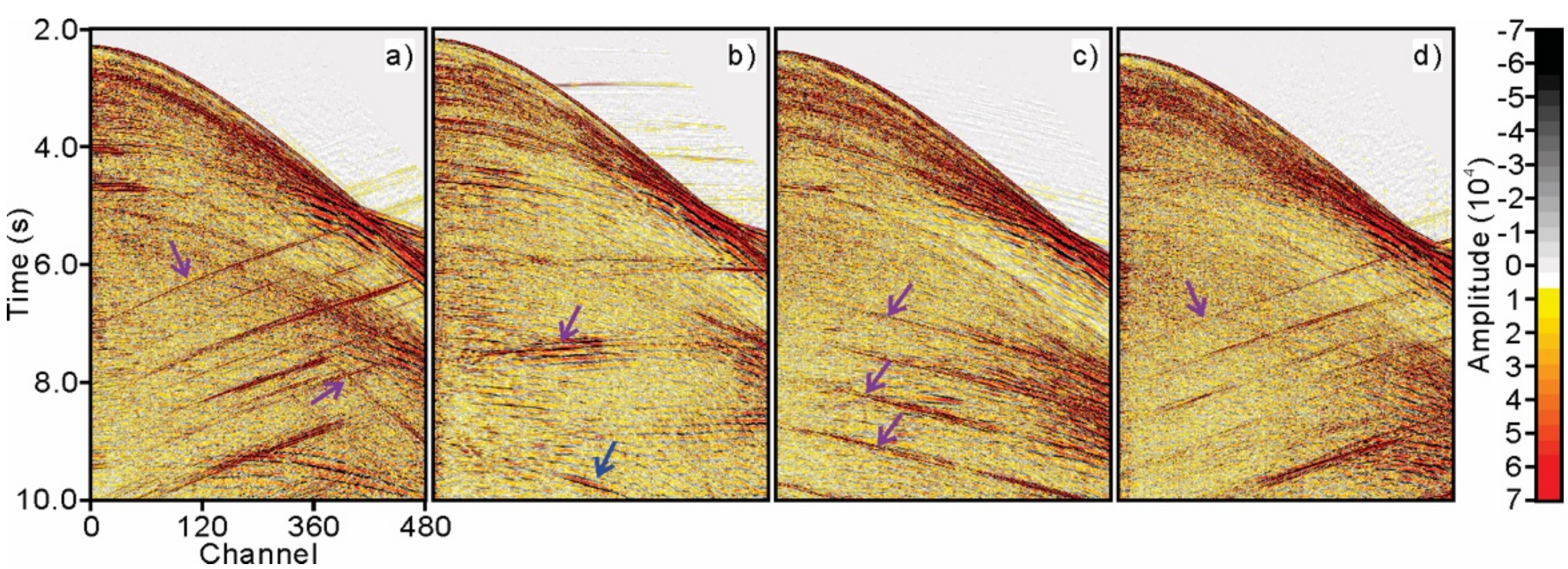

Figure 4: (a) to (d) show four shot gathers contaminated by different SI during the acquisition.

The purple arrows highlight SI noise. The blue arrow indicates a desired signal.

## THE EMPLOYED DNN-BASED WORKFLOW AND GEOPHYSICAL ALGORITHM

Figure 5 illustrates how the DNN-based workflow proposed by Sun et al. (2023) and Hou and Sun (2023) is applied to attenuate SI in the Camie field. To train the DNN, we constructed the training set as follows: First, 5 sail lines × 6 cables, covering all types of SI mentioned earlier, were selected and passed through a conventional geophysical SI attenuation algorithm. These accounted for 6% of the entire field SI-contaminated data from this survey.

The conventional geophysical algorithm involved preconditioning of common shot gathers by separation of dips of primary signal and SI noise where possible, and rank-reduction (Trickett et al., 2012) based denoising to improve signal-to-noise ratio (S/N) of coherent events (both wanted signal and unwanted SI noise). Next, SI noise was modelled by a sparse tau-p inversion (Zhang and Wang, 2015) scheme and adapted to the input common shot gathers with a complex wavelet adaptation which combines complex Morlet wavelet frame with unary complex Wiener filters (Ventosa et al., 2012). From this step, we obtained an estimate of the various types of SI (from now on called the SI model).

Our next step was to shuffle the SI model obtained from the conventional algorithm and blend it with SI-free shots. This survey has data of 5 sail lines × 3 cables that were naturally free from SI contamination during the acquisition. Data augmentation, including vertical translation and amplitude modulation, was applied to the SI-free shots and the SI model in order to generate more training pairs.

In addition, two procedures for improving signal fidelity were adopted. One was to inject a set of uniform random noise into the SI-contaminated input shot gather and the ground truth of the SI-free shot gather. The RMS amplitude of the random noise randomly varied from 8000 to 30000 for each shot gather. For context, the RMS amplitude of each field SI-contaminated shot in this survey typically ranged from 35000 to 45000. The other procedure was to use two adjacent shots on both sides of the to-be-processed shot were used as additional channels of the input. The DNN was trained to predict both the SI and a blend of the SI-free shot with the intact random noise. For a more detailed explanation of the reasoning behind these procedures, we refer readers to Sun et al. (2023) and Sun and Hou (2022).

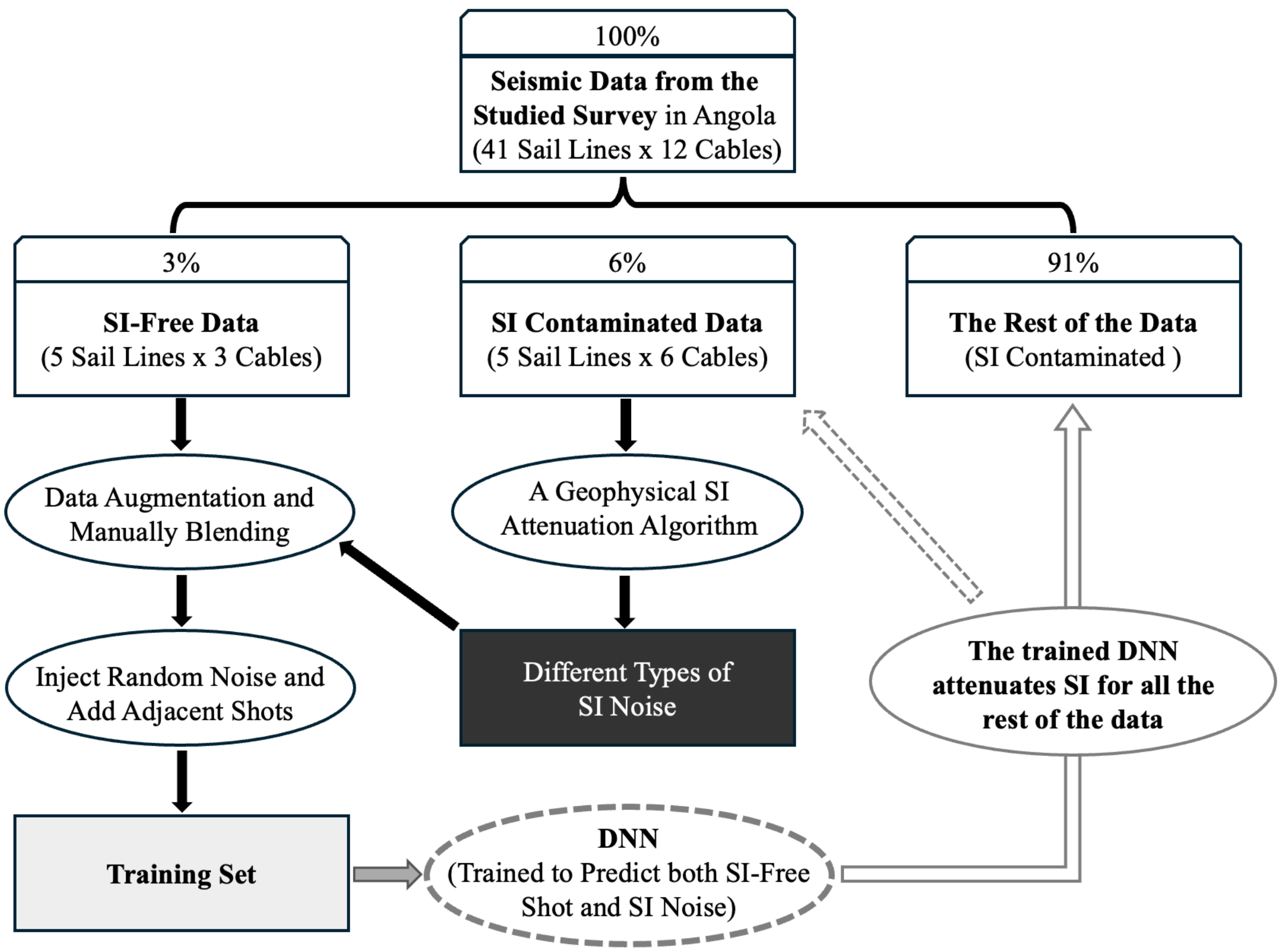

Figure 5: Illustration of how the DNN-based workflow proposed by Sun et al., (2023) is implemented in this survey.

Once the DNN is trained, it can be used to attenuate SI noise in the remaining SI-contaminated data from the entire survey (i.e., 91% of the dataset in this case study). In Figure 5, a dashed grey arrow links the trained DNN to the 6% SI-contaminated data that provided the SI model used in generating the training pairs. This is included to reflect that, in a commercial production setting, reprocessing this 6% using the trained DNN could be justified, as processing accuracy would be the top priority. However, from a purely research-oriented perspective, applying the trained DNN to data that contributed to the training process raises concerns regarding evaluation fairness. More detailed insights can be found in the discussion.

We employed the same DNN architecture as Sun et al. (2023), the study that introduced the DNN-based workflow. Their model was a slightly modified version of the original U-Net by Ronneberger et al. (2015). A visualization of the employed architecture can be found in Figure 6. Similar DNNs have been widely used in seismic data processing for various tasks.

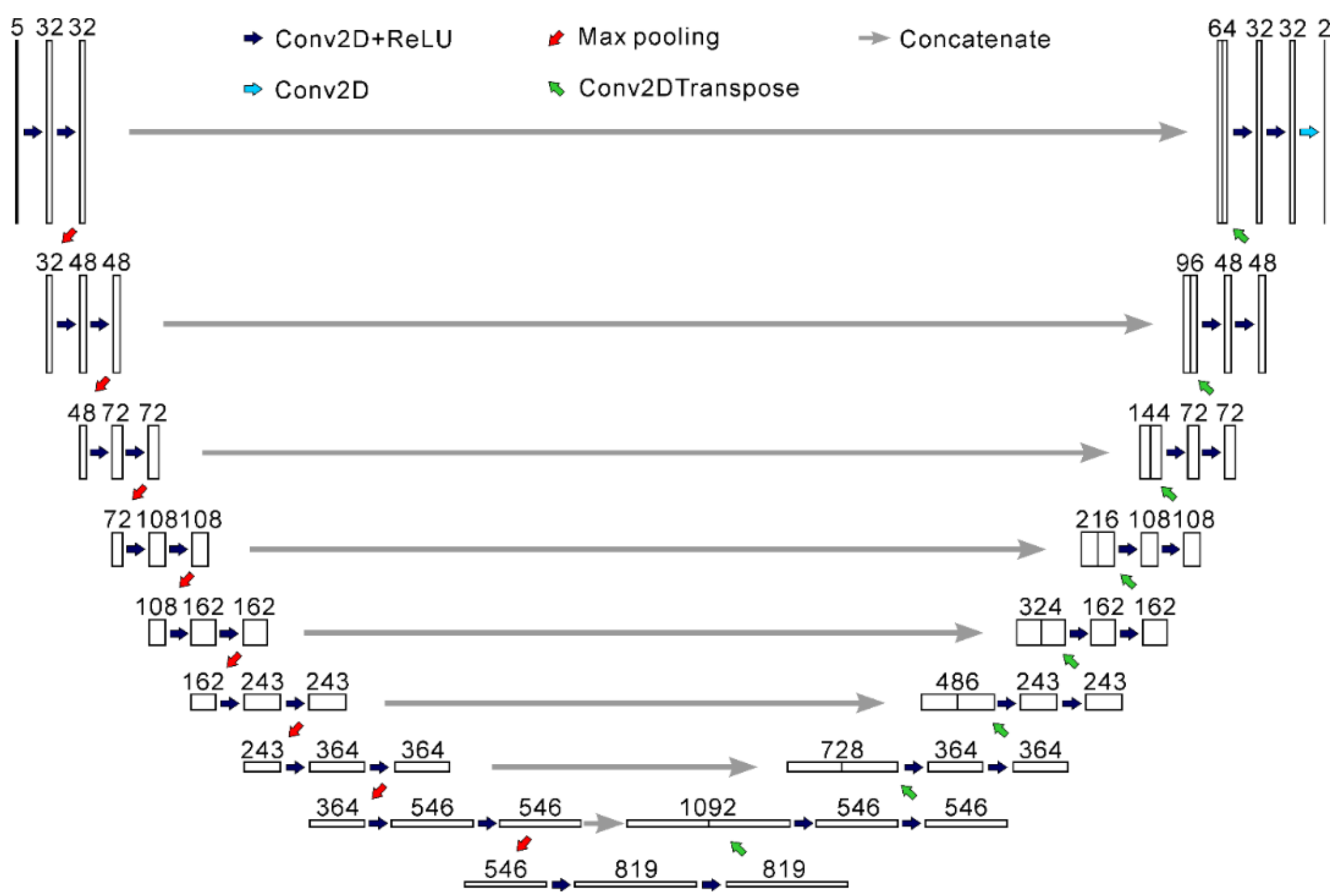

Figure 6: Architecture of the employed DNN (Sun et al., 2023). Each rectangle represents a collection of feature maps, and the numbers above (e.g. 32, 48, 72 and 108) represent the number

of feature maps.

The utilized DNN follows the standard U-Net structure with an encoder-decoder layout and skip connections that preserve spatial information by concatenating encoder features with their corresponding decoder layers. The encoder consists of 2D convolutional layers (filter size of 3×3) with ReLU (Rectified Linear Unit) activation, followed by max pooling layers (stride of 2 and a pool size of 2×2) that progressively downsample the input while extracting hierarchical features. The decoder uses transposed 2D convolutions to upsample the feature maps and reconstruct the output, with each level refining spatial details.

The main difference from the original U-Net is that, while the original U-Net doubles the number of feature channels (×2) at each downsampling step, this architecture increases them more gradually, by a factor of 1.5×, balancing model complexity and memory efficiency. No activation function is applied at the final output layer. After hyperparameter tuning, the Adam optimizer (Kingma and Ba, 2014), with a learning rate of 0.001, was selected, with a batch size of 4. The training process converged after approximately 80 epochs.

The mathematical formulations of the basic operations employed in this architecture—including 2D convolution, max pooling, transposed convolution, and concatenation—can be found in standard deep-learning literature such as Goodfellow et al. (2016), Dumoulin and Visin (2016), and the original U-Net paper by Ronneberger et al. (2015). This network architecture can be reproduced by using the original U-Net implementation (Ronneberger et al., 2015) and adjusting the number of feature maps according to those shown in Figure 6.

## RESULTS

In this section, we demonstrate the performance of the trained U-Net when applied to unseen data from the processed survey, based on a comparison with the conventional geophysical algorithm. We analyze the results in both the shot domain and the stacked CMP domain. Notably, although these examples are not drawn from the exact same cables (considering that the same dataset should never be repeatedly used as both the training and test datasets for a DNN), they serve as indicative representations of the quality of the SI model used to create the DNN's training inputs of the employed U-Net.

In addition to evaluating processing quality, we compare the processing time of the DNN-based workflow (employing the U-Net in Figure 6) with the case of solely using the conventional geophysical algorithm to process the entire survey.

**Comparison in Processing Quality**

As previously introduced, the field seismic shot gathers acquired from this survey are mainly contaminated by four types of SI. To provide clarity, examples of each type are shown in Figures 4a to 4d. In this section, we show the SI attenuation results obtained by employing the conventional geophysical algorithm in Figures 7a to 7d. In these results, visible SI residuals are evident in each example, as denoted by the purple arrows. Additionally, some signal damage can be observed, as indicated by the blue arrows in Figure 7b and Figure 4b.

For comparison, the SI attenuation results from the DNN-based workflow (employing the U-Net in Figure 6) are shown in Figures 8a to 8d. These results clearly show much better quality. The SI residuals are significantly reduced compared to those from the conventional algorithm in Figure 7a to 7d. This improvement is even more noticeable when comparing the removed SI. Figures 9a to 9d show the SI removed by the conventional geophysical algorithm, which is incomplete and discontinuous. In contrast, the DNN-based workflow produces much more continuous and accurate SI extraction, with notably less signal leakage, as shown in Figures 10a

to 10d.

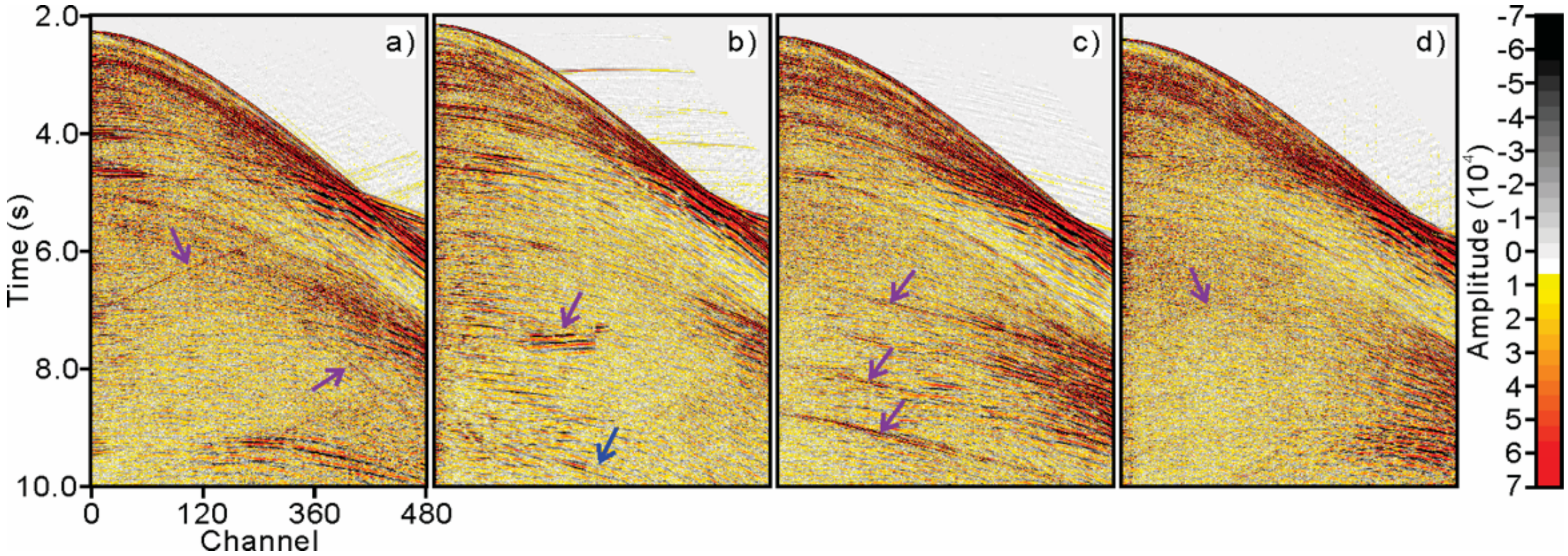

Figure 7: (a) to (d) show the SI attenuation results of the conventional algorithm for the four SI-contaminated shot gathers in Figure 4a to 4d. The purple arrows indicate the SI residual, while the blue arrow indicates the signal damage caused by the conventional geophysical algorithm.

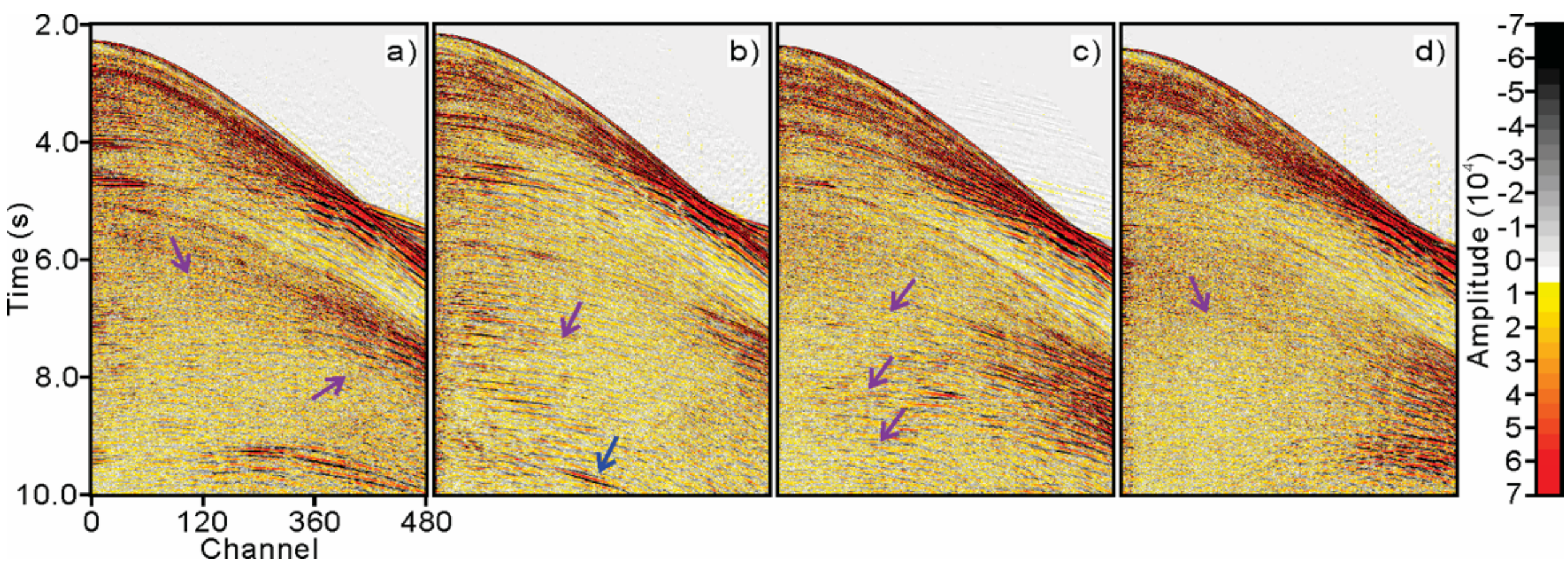

Figure 8: (a) to (d) show the SI attenuation results of the DNN-based workflow (employing the U-Net in Figure 6) for the four SI-contaminated shot gathers in Figure 4a to 4d. The purple arrows indicate much less SI residual compared to Figure 7a to 7d. The blue arrow indicates that less signal damage was caused by the trained U-Net than by the conventional geophysical algorithm (Figure 7b).

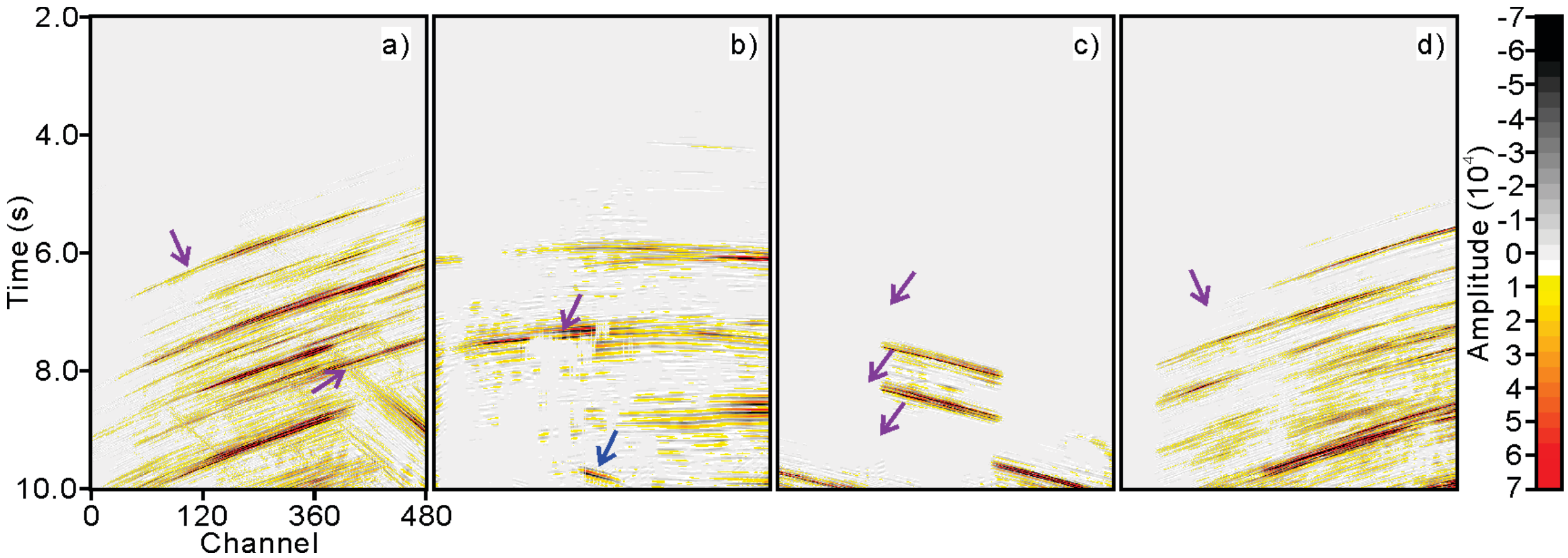

Figure 9: (a) to (d) show the attenuated SI of the conventional geophysical algorithm for the four SI-contaminated shot gathers in Figure 4a to 4d. The purple arrows indicate the incomplete SI removal, while the blue arrow indicates the signal leakage caused by the conventional algorithm.

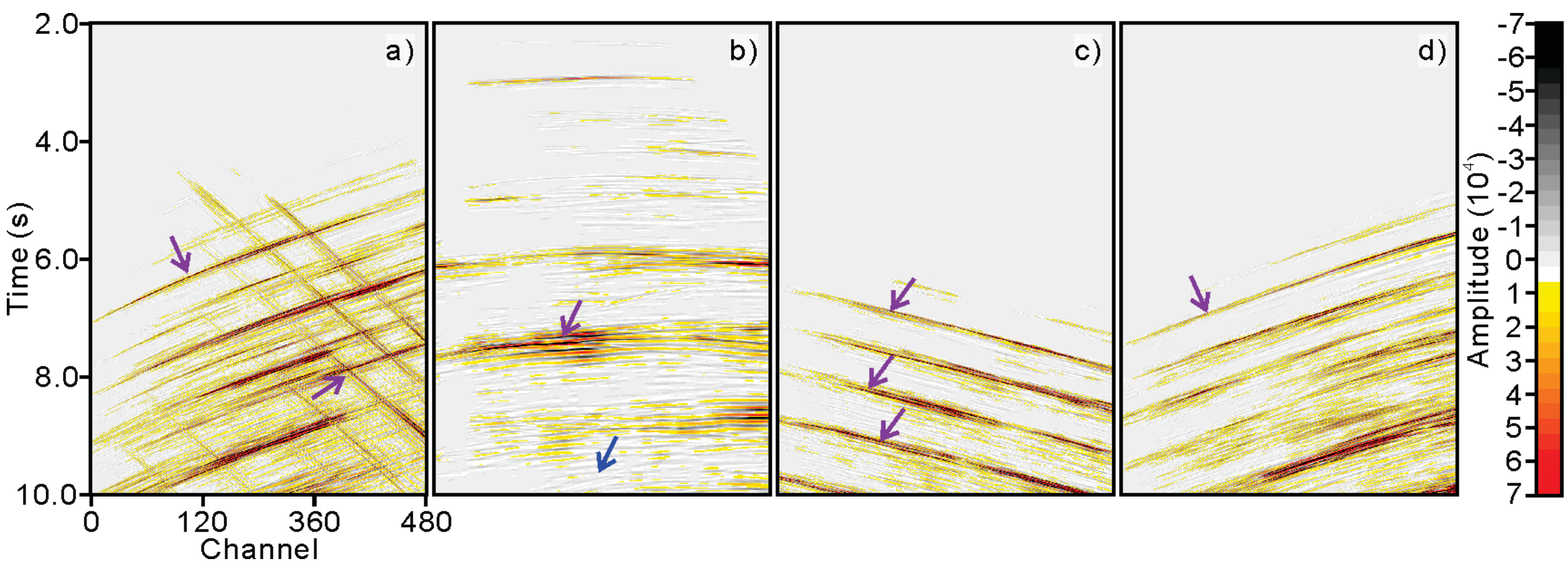

Figure 10: (a) to (d) show the attenuated SI of the DNN-based workflow (employing the U-Net in Figure 6) for the four SI-contaminated shot gathers in Figure 4a to 4d. The purple arrows indicate a much more complete SI removal compared to Figure 9a to 9d. The blue arrow indicates that almost no signal leakage was caused by the trained U-Net.

An example in the stacked CMP domain is given in Figure 11. Figure 11a shows a time window extracted from a CMP stack of data contaminated by SI during the acquisition. Figures 11b and 11c then show the SI attenuation results obtained using the conventional geophysical algorithm and the DNN-based workflow, respectively. While the overall difference is not easily noticeable, a relatively distinct difference is highlighted by the blue boxes in the bottom left.

For better comparison, Figures 11d and 11e show the removed SI using the conventional geophysical algorithm and the DNN-based workflow, respectively. As observed in the shot-domain examples, the DNN-based workflow demonstrates a significant improvement in processing quality compared to the conventional geophysical algorithm. It exhibits a more complete and comprehensive extraction of the SI, as indicated by the blue circle, and significantly reduces signal leakage, as indicated by the blue arrows.

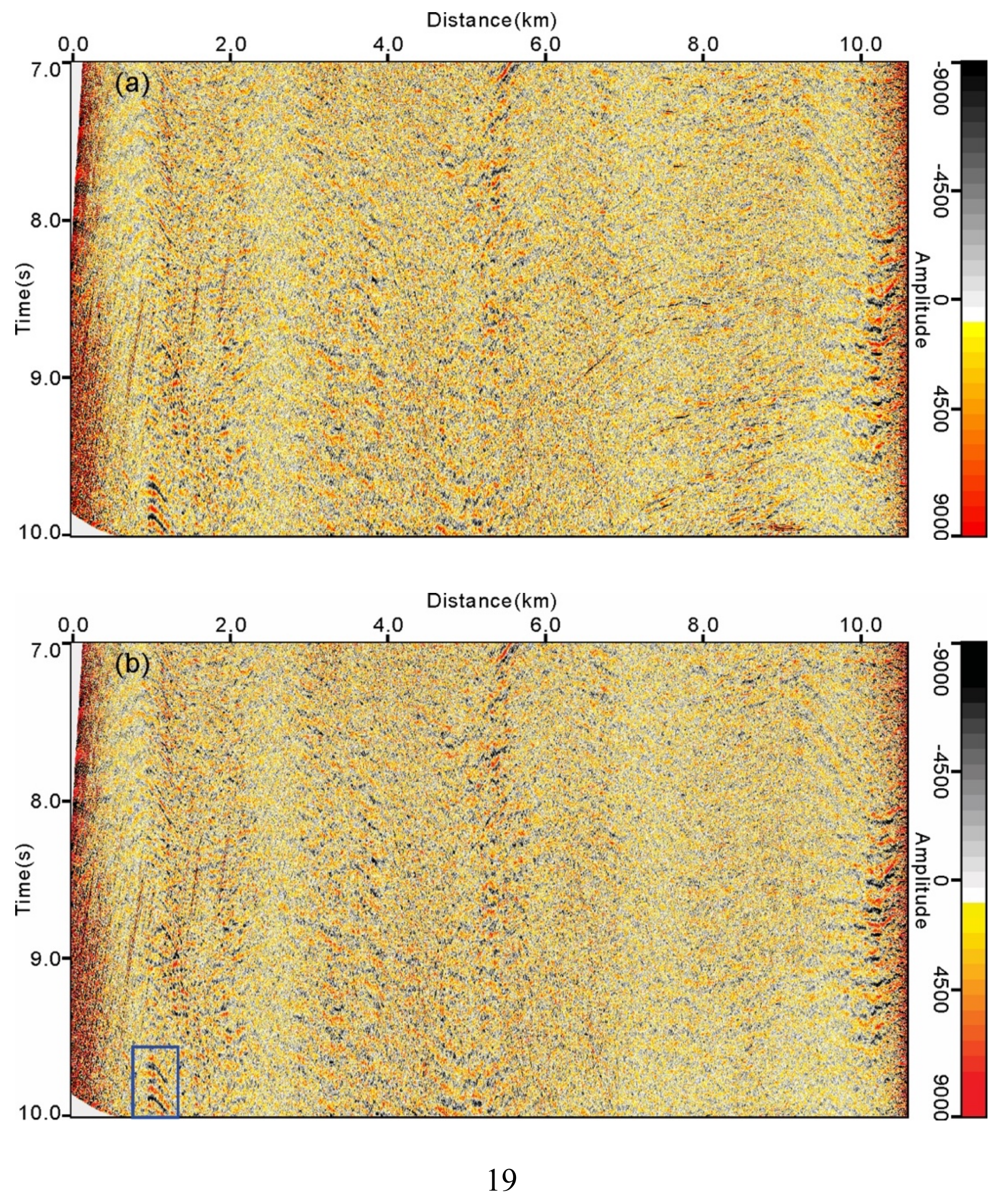

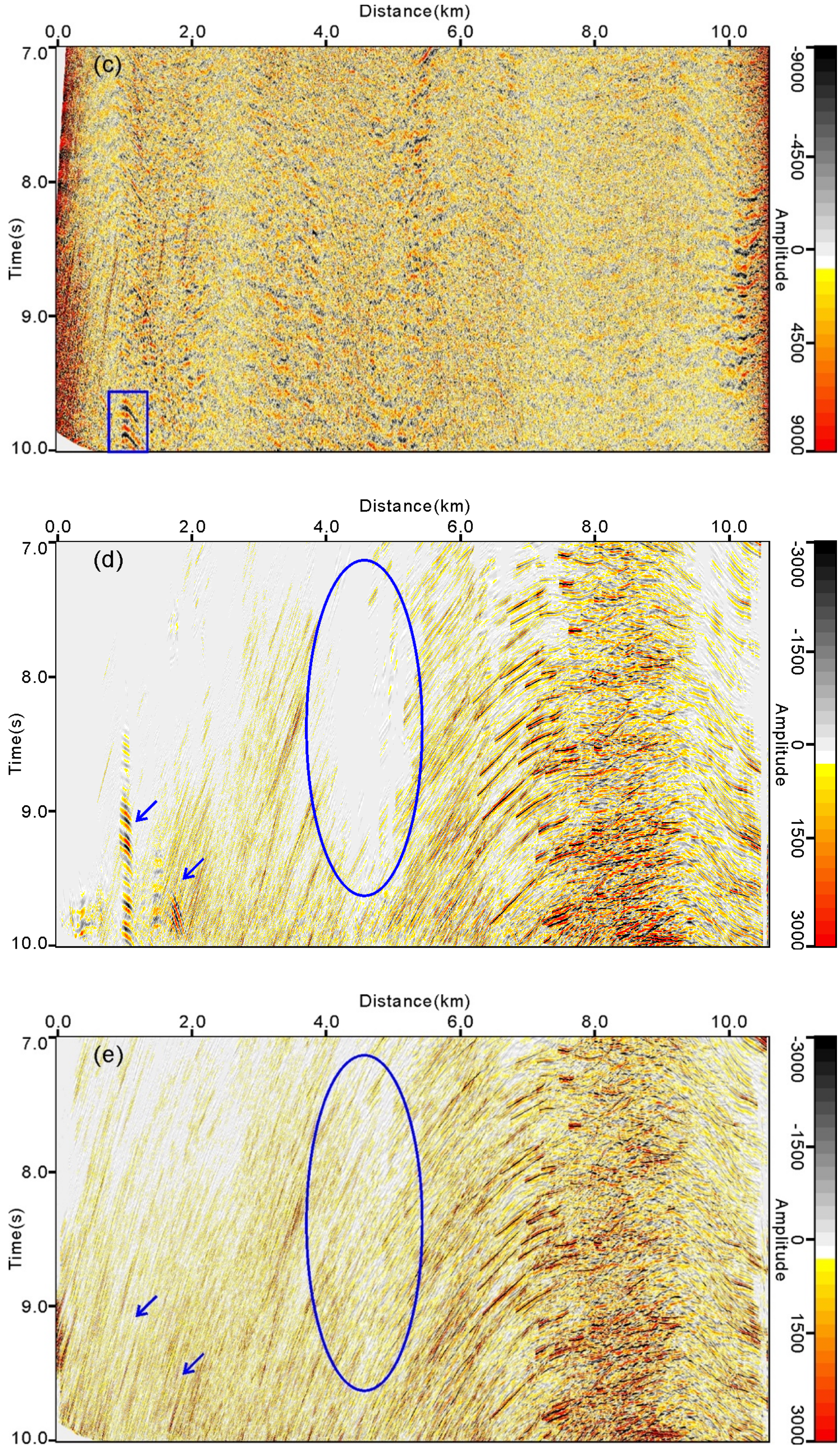
Distance(km)
0.0
2.0
4.0
6.0
8.0
10.0
Time(s)
7.0
8.0
9.0
10.0
(c)
-9000
-4500
0
4500
9000
Amplitude
(d)
-3000
-1500
0
1500
3000
(e)

Figure 11: CMP stacks of (a) the field data contaminated by SI during the acquisition, (b) SI attenuation result using the conventional geophysical algorithm, (c) SI attenuation result using the DNN-based workflow (employing the U-Net in Figure 6), (d) attenuated SI using the conventional geophysical algorithm and (e) attenuated SI using the DNN-based workflow (employing the U-Net in Figure 6). Arrows and ellipses respectively highlight the comparisons of signal leakage and the completeness of SI extraction between the two methods.

These observations are also supported by the comparison of the RMS amplitude, calculated based on the average RMS per shot, specifically in the late time (i.e., 8.0 s to 10.0 s) and high frequency (i.e., above 20 Hz) sections, where the SI mainly exists. Figure 12 presents the deep high-frequency RMS levels of two sail lines located at different areas of the survey. Each column shows examples for comparison from the same sail line: SI-contaminated raw data (top), SI attenuation result using the conventional geophysical algorithm (middle), and SI attenuation result using DNN-based workflow (bottom). In the raw data panels, intense vertical stripe patterns, appearing in red and yellow, highlight strong SI contamination. The conventional geophysical algorithm partially suppresses them, but residual striping and signal distortion, seen as light green streaks, remain. In contrast, the SI attenuation results from the DNN-based workflow show a notable reduction in SI, with the vertical artifacts largely removed and the background appearing more uniform in blue-green shades. This color transition illustrates improved continuity and preservation of coherent signals in the deeper channels.

Figure 13a illustrates the overall deep high-frequency RMS level for the entire survey. The comparison of the two workflows across the entire survey is shown in Figures 13b and 13c. In the raw SI-contaminated data (Figure 13a), extensive high RMS values appear throughout the survey, especially along central and northeastern sail lines, highlighting severe SI contamination, as indicated by the high-amplitude regions in red. The conventional geophysical algorithm (Figure 13b) achieves moderate attenuation, reducing SI in less affected regions; however, significant residuals remain in areas of heavy contamination. In contrast, the DNN-based workflow (Figure 13c) exhibits a more consistent and effective SI reduction across the entire survey. Most sail lines show lower RMS levels, transitioning into green and blue, indicating better suppression of SI energy even under stronger contamination conditions. It is also worth noting that, although the DNN-based workflow overall outperforms the conventional geophysical algorithm, both methods exhibit comparable limitations in the regions most heavily affected by SI contamination (e.g., central and northeastern sail lines).

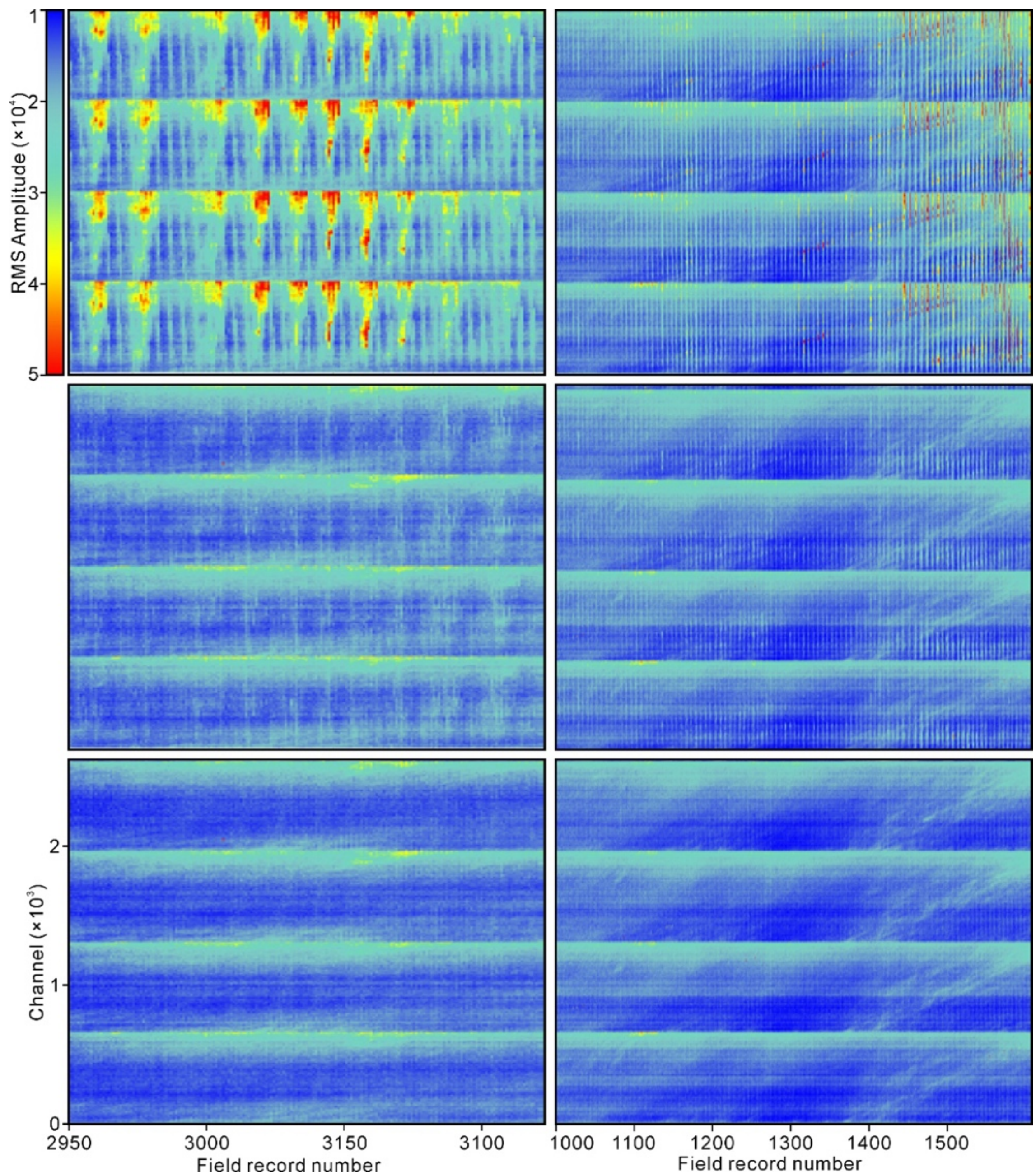

Figure 12: Deep high-frequency RMS of two sail lines in the studied survey. In each column the figures from top to bottom are respectively the SI-contaminated raw data, SI attenuation result using the conventional geophysical algorithm and SI attenuation result using the DNN-based workflow (employing the U-Net in Figure 6). SI appears predominantly as vertical stripes in these plots.

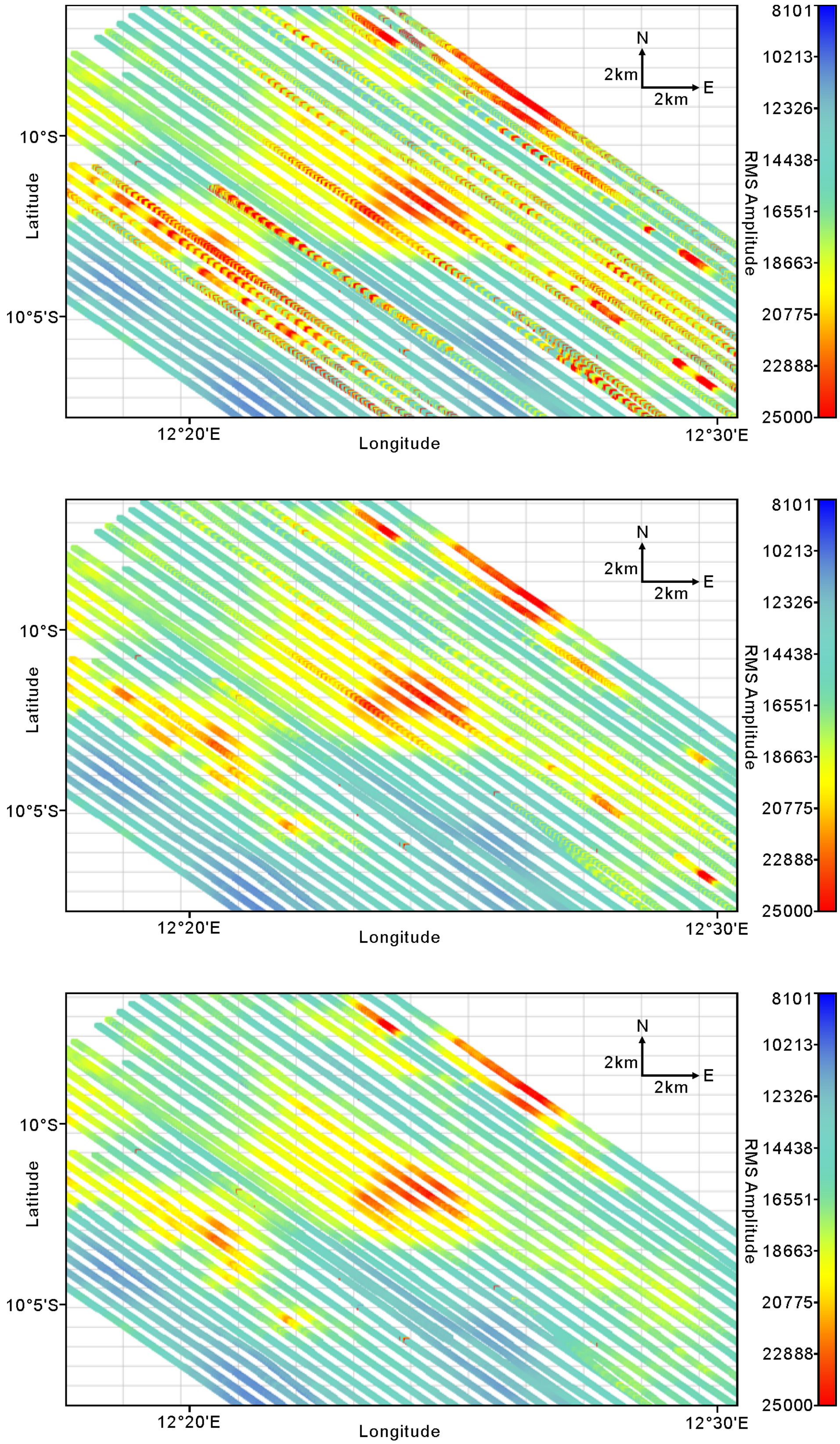

Figure 13: Deep high-frequency RMS of the studied survey. (a), (b) and (c) represent

respectively the SI-contaminated raw data, SI attenuation result of the conventional geophysical algorithm and SI attenuation result of the DNN-based workflow (employing the U-Net in Figure 6). Each slant line represents a sail line of the survey, and the color indicates the RMS amplitude.

**Comparison in Processing Time**

Processing one sail line with the conventional geophysical algorithm took approximately 3.5 hours, excluding the time spent on parameter optimization. When scaled up for the entire survey, which consists of 41 sail lines in total, the computational time would amount to approximately 143.5 hours using a single GPU (Nvidia Quadro RTX 6000). In the DNN-based workflow, data from 6 cables × 5 sail lines were first processed using the conventional geophysical algorithm to obtain SI models of different SI types, which took 8.75 hours. Following this, the data preparation and the DNN training process took approximately 35 hours.

Once the DNN was trained, it could process each sail line in just 15 minutes, amounting to 10.25 hours for the entire survey. Note that when computing this time cost, we take into account the potential reprocessing of the SI-contaminated sail line that provided the SI model used in the DNN's training pairs. As a result, the total computational time for the DNN-based workflow was approximately 54 hours using the same computational resources, which is roughly one-third of the time required by the conventional geophysical algorithm. The total area of the studied survey is 345 $km^2$. The number of shot gathers used here has met the needs of DNN training. If applying the same DNN-based workflow to even larger surveys, the DNN training time will not increase significantly, but the efficiency benefits of the DNN during the application phase will be further scaled up.

## DISCUSSION

Deep learning-based methods are recognized for their great potential but have yet to be widely implemented in real-world seismic processing projects. One key reason is that their signal fidelity often falls short of industry standards. To address this problem, much research has been conducted. From the promising results shown in our presented case study, we see positive signs that deep learning could benefit some real-world tasks; at least, it has started contributing to our focused area of SI attenuation.

We provide a detailed analysis of the effectiveness of a DNN-based workflow based on a comparison with a conventional geophysical algorithm. The data used for this study was acquired from a real-world marine seismic survey block off the coast of Angola, covering over 345 $km^2$. The DNN-based workflow demonstrated greater overall processing time than the conventional geophysical algorithm. Specifically, the entire process, including data preparation and DNN training, required less than half the computational time of the conventional geophysical algorithm. This efficiency gain is expected to scale with the increase in the size of the processing area.

Apart from processing time, the DNN-based workflow showed a general improvement in processing quality over the conventional geophysical algorithm, especially in reducing SI in regions with poor SnR or where the SI noise has similar dips to the desired signal. Our field data examples demonstrate that even the DNN model was trained on an incomplete and discontinuous SI model produced from the conventional workflow, it could effectively attenuated SI during the inference phase.

These results mainly from two factors: Firstly, although not perfect, the SI models derived from the conventional algorithm cover the types of SI that the DNN model is expected to learn to address. For SIs that are very weak in amplitude and cannot be distinguished from the signal using the conventional geophysical algorithm, the DNN model would also struggle to detect and identify them due to their absence during the training phase. Secondly, there are some data truly free from SI during the acquisition of this survey, which can serve as satisfactory-quality ground truth of the desired signal, specifically, 5 sail lines × 3 cables out of the total 41 sail lines × 12 cables.

A more challenging scenario arises when the entire survey is contaminated by SI, forcing us to use SI-attenuated shots—also produced from a conventional geophysical algorithm—to create the training pairs for the DNN. In such cases, refining the conventional geophysical algorithm becomes essential, as it plays a key role in producing both the SI model and the pseudo-SI-free shots needed for DNN training. This introduces some uncertainties regarding the performance of the DNN-based workflow.

In this study, all displayed examples are from applying the trained DNN to unseen SI-contaminated data—specifically, the 91% of the raw dataset not involved in training—and are compared against SI attenuation results from the conventional geophysical algorithm. From a practical standpoint in a production setting, where the primary objective is to enhance processing quality, applying the trained DNN to that 6% of raw SI-contaminated data (used to generate the SI models for training pair construction) could also be justified. However, from a methodological perspective, doing so raises concerns about potential data leakage and evaluation fairness. Although the DNN was not trained on the SI-free versions of these shots (as such versions do not naturally exist), the SI models were derived from them, introducing a partial dependency. To maintain a clear separation between training and inference, we intentionally excluded this 6% from the evaluation examples, though we acknowledge it as a potentially reasonable step in real-world commercial workflows.

In addition, several limitations must be acknowledged. One critical concern is overfitting, particularly when the DNN is exposed to limited or homogeneously structured training data. While we mitigated this risk through diverse augmentation and pairing strategies, it remains a concern. A separate but related concern arises if the training SI models do not adequately reflect the full complexity present in the field survey. This also explains why, in the present case study, we chose to process 5 sail lines × 6 cables, rather than all cables from only one or two sail lines. Although both approaches would yield a similar number of SI models, the former ensures broader coverage of the SI types encountered across the survey. In practice, this places a higher demand on the geophysicist operating the conventional geophysical algorithm, as they must handle a wider range of SI characteristics from different sail lines.

Moreover, the employed DNN-based workflow is a purely data-driven method, and its reliance on survey-specific training may limit its generalizability to other seismic surveys. Our experience suggests that a DNN model trained on one survey is unlikely to perform well on a different survey without retraining or fine-tuning. This reflects a domain shift challenge, as the DNN struggles to adapt to new data distributions not represented in the training set. This underscores the potential need for domain adaptation strategies before broader deployment.

While these cases are beyond the scope of our present study, we recognize their significance and advocate for further exploration. Further investigation into the robustness and capabilities of this DNN-based workflow is necessary. Future work could also include exploring the application of this DNN-based workflow to other coherent noise types in seismic processing.

## CONCLUSIONS

We applied a DNN-based workflow to attenuate SI for an entire marine seismic survey block in the Camie Field of Angola, covering over 345 $km^2$. Specifically, a U-Net was employed. This area is marked by various types of SI and significant maritime activity. We provide a thorough comparison of the DNN-based workflow versus a conventional geophysical algorithm on shot gathers, CMP stacks, and deep high-frequency RMS analysis. The DNN-based workflow outperforms in both processing quality and time. This paper presents the largest-ever case study comprehensively comparing a DNN-based workflow with a conventional geophysical algorithm in a real-world SI attenuation project.

Despite this promising case study, further investigation of this DNN-based workflow's robustness and capability is needed. Up to now, the implementation of such a DNN-based workflow still relies on a geophysical algorithm to provide at least the labels for the SI, and those labels are expected to cover all the SI types that the DNN is expected to process at the inference stage. A DNN under supervised learning is a purely data-driven technique for which the quality of the training pairs is crucial; thus, developing better geophysical algorithms remains important. The selection of good data samples also requires human effort and experience.

Overall, for large-scale real-world processing, it is still essential to begin with a solid conventional geophysical algorithm, as it forms the foundation upon which the DNN can provide further improved SI-attenuation results. Ultimately, only processors with experience in both geophysical and deep learning methods will be able to balance the effort needed to tune these two parts and achieve output of the best value. Besides, future work may include applying the DNN-based workflow to other surveys, evaluating alternative DNN architectures, and exploring the incorporation of physics-based constraints to enhance generalization.

## ACKNOWLEDGMENTS

The authors thank VIRIDIEN (formerly CGG) Earth Data for granting permission to use the field seismic data from Angola and the location map for the purposes of research and publication. The authors thank Dr. Vetle Vinje for his invaluable insights. The authors thank Mr. Cian Ryan and Mr. Harrison Moore for their contributions to the data processing using the geophysical algorithm employed as a benchmark in this study. This research work was funded by VIRIDIEN and NFR (Norwegian Research Council) through an industrial Ph.D. grant (project number 314179). The first author thanks the University of Oslo and VIRIDIEN (formerly CGG) for providing the environment and support in which this research was carried out. This work was completed while the first author was affiliated with the University of Oslo and VIRIDIEN (formerly CGG).